\documentclass{appolb}
\usepackage{amsmath}
\usepackage{bbm}
\usepackage{graphicx}
\usepackage{psfrag}
\begin{document}

\title{From 4D Reduced SYM Integrals \\
to Branched-Polymers }
\author{Zdzislaw Burda$^1$, Bengt Petersson$^2$ and Marc Wattenberg$^2$
\address{
$^1$ Institute of Physics, Jagellonian University\\
ul. Reymonta 4, 30-059 Krakow, Poland\\[0.4cm]
$^2$ Fakult\"at f\"ur Physik, Universit\"at Bielefeld \\
P.O. Box 100131, D-33501 Bielefeld, Germany \\[0.4cm]}}
\maketitle

\begin{abstract}
\begin{center}
We derive analytically one-loop corrections 
to the effective\\ Polyakov-line operator 
in the branched-polymer approximation of the\\ reduced
four-dimensional supersymmetric Yang-Mills integrals.
\end{center}
\end{abstract}

\section{Introduction}
The IKKT matrix model, called also IIB matrix model,
was proposed \cite{ikkt} 
as a candidate for a non-perturbative formulation of 
superstring theory. Its partition function is defined 
by reduced ten-dimensional supersymmetric 
Yang-Mills integrals, which generate a quantum 
extension of the sum over topologies of 
the string world-sheet. In the large $N$
limit the action of the IKKT model approaches the action
of type IIB superstrings. The one-loop approximation of the model can be 
rewritten as a geometrical model of graphs \cite{ss}.
The vertices of these graphs, which are dressed with ten-dimensional vectors,
contain information about the space-time location 
of the world-sheet points of the corresponding string. 
The graphs are closely 
related to branched-polymers which are known to have fractal 
dimension equal four. This means that the scaling
of physical observables within the ten-dimensional 
model of graphs is expected to be 
given by that of the four-dimensional theory \cite{ss,mm}.
This fact is treated as an indication of the spontanous
symmetry breaking of the ten-dimensional Lorentz symmetry
to its four-dimensional subgroup. 

Counterparts of the IKKT model can be defined in four and six
dimensions. In four dimensions, the one-loop 
approximation simplifies, the corresponding
graph model reduces to a model 
of tree graphs called also branched-polymers. 
Contrary to the corresponding graphs 
in six and ten dimensions, where the integration 
over fermionic degrees of freedom leads to a sign problem, each 
graph in the branched-polymer ensemble of the four-dimensional
model has a positive definite weight.
Therefore the four-dimensional model is frequently treated
as a test-bed for the IKKT model. 
In fact many features of the model were first discovered in 
four dimensions, where one could explicitly 
carry out calculations \cite{kps,b1,b2,a1,a2,a3}.\\
\\
 In this paper we continue the studies of the 
four-dimensional case. We address the problem of the 
approximation of operators. In particular we derive the 
form of the Polyakov-line operator for the corresponding
branched-polymer model.

\section{One-loop approximation}
The partition function of the 4D IIB matrix model
is given by a supersymmetric Yang-Mills reduced integral \cite{ikkt}~:
\begin{equation}
Z=\int{\cal{D}}A{\cal{D}}\bar\Psi{\cal{D}}\Psi\
e^{-{S}[A,\bar\Psi,\Psi]}
\label{Z} 
\end{equation}
with the action
\begin{equation}
S[A,\bar\Psi,\Psi]=-\frac{1}{4g^{2}}
{\rm Tr}\;[A^{\mu},A^{\nu}]^{2}-\frac{1}{2g^{2}}
{\rm Tr}\;\bar\Psi\Gamma_{\mu}[A^{\mu},\Psi]\quad,
\label{S} 
\end{equation}
where $A^{\mu}(\mu=1,...,4)$ 
are traceless $N\times N$ Hermitian matrices and
$\bar\Psi,\Psi$ are $N\times N$ traceless matrices of Grassmann variables, which 
transform as Weyl-spinors. 
For convenience the spinor indices of $\bar\Psi^a,\Psi^a$, $a=1,2$ and matrix indices $ij$ of
the matrices $A^\mu_{ij}$, $\bar{\Psi}^a_{ij}$ and $\Psi^a_{ij}$ are suppressed in (\ref{Z},\ref{S}).
The measures of integration are flat in the space of traceless
matrices~:
\begin{eqnarray*}
{\cal{D}}A&\equiv&\prod_{\mu=1}^{4}
\left(\prod_{i=1}^{N} {\rm d} A^\mu_{ii}\ 
\delta\left(\sum_{i=1}^{N} A^\mu_{ii}\right)
\prod_{i<j} {\rm d} {\rm Re}A^{\mu}_{ij}\ 
{\rm d} {\rm Im}A^\mu_{ij}\right)\\
{\cal{D}}\bar\Psi{\cal{D}}\Psi&\equiv&\prod_{a=1}^{2}\left(
\prod_{i=1}^{N}{\rm d}\bar\Psi^a_{ii}\
{\rm d}\Psi^a_{ii}\ 
\delta\left(\sum_{i=1}^{N}\bar\Psi^a_{ii}\right)\ 
\delta\left(\sum_{i=1}^{N}\Psi^a_{ii}\right) \right) \\ &&
\prod_{a=1}^2 \left(
\prod_{i<j}{\rm d} {\rm Re} \bar\Psi^a_{ij}\
{\rm d}{\rm Re}\Psi^a_{ij}\ {\rm d}{\rm Im} \bar\Psi^a_{ij}\ {\rm d}{\rm Im}
\Psi^a_{ij}\right) \quad .
\end{eqnarray*}
The model is invariant with respect to Lorentz transformations.
It has also ${\cal N}\!=\!2$ supersymmetry and 
$SU(N)$ invariance, being a remnant
of the gauge invariance of 
the full non-reduced model.

Since the model has not yet been solved analytically, 
one way of getting insight into
its behaviour is to use approximate methods such as, for example,
the one-loop approximation \cite{ss,mm}. 
The idea behind the one-loop approximation is to split the fields 
$A^\mu$, $\bar{\Psi}$ and $\Psi$ into the classical part and 
quantum fluctuations and further to integrate out 
the quantum fluctuations using the Gaussian approximation.
In this way one derives an effective
one-loop action, which depends only on the classical fields. 
One can see that a classical vacuum for the action 
(\ref{S}) is given by a set of diagonal matrices 
$A^\mu$, $\bar{\Psi}$ and $\bar{\Psi}$,
or by any set of matrices, which can be obtained 
from it by a symmetry transformation. 
The one-loop approximation was worked out in  paper \cite{ss}.
Starting point of the approximation is to devide
the matrices into the diagonal 
(classical) and off-diagonal (quantum) parts~:
\begin{eqnarray}
\label{xa}
A^{\mu}_{ij}&=&x^{\mu}_i\delta_{ij}+a^{\mu}_{ij}\\[0.5cm]
\bar\Psi^a_{ij}&=&\bar\xi^a_{i}\delta_{ij}+\bar\psi^a_{ij}
\nonumber\\[0.5cm]
\Psi^a_{ij}&=&\xi^a_{i}\delta_{ij}+\psi^a_{ij}\quad,
\nonumber
\end{eqnarray}
where $\sum_{i}x^{\mu}_{i}=0$ and
$\sum_{i}\bar\xi^a_{i}=0$\ $\left(\sum_{i}\xi^a_{i}=0\right)$,
respectively.
Let us denote the one-loop effective action by 
$S_1[x,\bar{\xi},\xi]$. The partition function (\ref{Z})
is approximated by~:
\begin{equation*}
Z_1 = \int {\cal D} x {\cal D} \bar{\xi} {\cal D} \xi
e^{-S_1[x,\bar{\xi},\xi]} \quad .
\end{equation*}

For large separations between the eigenvalues 
$|x^{\mu}_i -x^{\mu}_k| \gg \sqrt{g}$ the effective one-loop action
can be shown to admit the following form~:
\begin{equation}
S_1[x,\bar{\xi},\xi] = 
-\frac{1}{2} \sum_{i<j} {\rm tr} \; C^2_{ij}\quad,
\label{S1L}
\end{equation}
where the trace ${\rm tr}$ is taken in the Lorentz indices of $C_{ij}^{\mu\nu}$ \begin{eqnarray}
C^{\mu\nu}_{ij}&\equiv&
(\bar\xi_{i}-\bar\xi_{j})\Gamma^{\mu\sigma\nu}(\xi_{i}-\xi_{j})
\frac{(x^{i}_{\sigma}-x^{j}_{\sigma})}{||x^{i}-x^{j}||^{4}}\quad.
\label{C}
\end{eqnarray}
The symbol $|| . ||$ denotes the length of a vector~:
$|| x_i-x_j ||=\sqrt{\sum_\mu(x^\mu_i-x^\mu_j)^2}$ and
$\Gamma^{\mu\sigma\nu} =
\frac{1}{3!}\Gamma^{[\mu}\Gamma^{\sigma}\Gamma^{\nu]}$.
In four dimensions this symbol can be simplified to
$\Gamma^{\mu\sigma\nu} = 
\epsilon^{\mu\sigma\nu\tau} \Gamma_\tau \Gamma_5$.

One can further simplify the problem by the explicit integration
of the remaining fermionic zero modes $\bar{\xi},\xi$.  
For large separations between the eigenvalues $x^\mu_i$ one obtains~:
\begin{equation}
\int {\cal D} \bar{\xi} {\cal D} \xi \;
e^{-S_1[x,\bar{\xi},\xi]}
= \sum_T \prod_{(ij)\in T} ||x_i-x_j||^{-6}\quad,
\end{equation}
where the sum runs over all tree graphs consisting of 
$N$ vertices and $(N\!-\!1)$ links.  
Each vertex is dressed with a vector $x_i^\mu$, $\mu=1,\dots, 4$. 
The product runs over all links $(ij)$ of the tree graph, giving
the total weight, which is equal to a product
of link-weights depending on the link length.

If the eigenvalues come close to each other, one
has to introduce corrections to the approximation \cite{ss,mm}.
One can explicitly calculate within the matrix model,
that a strong repulsion appears in the
effective potential for $||x_i-x_j|| < \sqrt{g}$, which 
prevents the eigenvalues from approaching each other. Since one 
is interested in the large distance behaviour of 
the model, the details of the model
on short distances are not important.
One can therefore model this repulsion by
adding a hard core to the potential, which keeps each
pair of eigenvalues away from each other. Having done this,
one eventually arrives at the partition function \cite{ss}~:
\begin{equation*}
Z_{bp} = \sum_T \int \prod_i{}' {\rm d}^4 x_i 
\left(\prod_{ab} \theta\left(||x_a-x_b||-c\sqrt{g}\right)\right)
\prod_{(ij)\in T} ||x_i-x_j||^{-6}\quad,
\end{equation*}
which describes an ensemble of branched-polymers embedded
in four dimensions with a hard core. The core is modelled 
by the Heaviside step function $\theta(x)$
and prevents any two vertices from being closer 
to each other than the core size, which is proportional to
$\sqrt{g}$ with some irrelevant constant of proportionality $c$.
As above the sum runs over all trees. The integral is taken
over all positions of tree vertices, such that the
center of mass of the polymer is at zero~:
$\delta^4(\sum_j x_j)$, which is indicated as a prime at the product
in the integration measure. This constraint is a remnant of the
tracelessness of the matrices. It removes the translational
zero mode of the integral, which would otherwise make the integral
divergent. The product over $ab$ runs over all pairs of vertices, 
while the one over $(ij)$ only over pairs of vertices joined by 
a link of the tree graph $T$. One can write the partition function as follows~:
\begin{equation}
Z_{bp} = \sum_T \int \prod_i{}' {\rm d}^4 x_i \ e^{-S_{bp}[T,x]} 
\end{equation}
with the action
\begin{equation}
S_{bp}[T,x] = 6 \sum_{(ij)\epsilon T} \ln || x_i - x_i || 
+ H \sum_{ab} \theta\left(c\sqrt{g} -||x_a-x_b||\right)\quad,
\label{s1bp}
\end{equation}
where $H$ is a huge positive constant $H\rightarrow\infty$, which 
takes care of the short distance repulsion between 
any two points.
The configuration space changes in the course of 
approximating the theory. At the beginning, it is given
by $N\times N$ matrices $(A, \bar{\Psi},\Psi)$, 
then by diagonal matrices $(x,\bar{\xi},\xi)$
and eventually by branched-polymers dressed with 
position vectors $(T,x)$. The partition function and the effective
action change correspondingly~: 
$$Z \rightarrow Z_1 \rightarrow Z_{bp} \ ,$$
$$S[A^\mu,\bar{\Psi},\Psi] \rightarrow 
S_1[x^\mu,\bar{\xi},\xi] \rightarrow S_{bp}[T,x^\mu] \ .$$ 
To complete the approximation
scheme, one has also to find out how the operators change.
For any operator ${\cal{O}}[A^\mu,\bar{\Psi},\Psi]$ defined 
in the original theory, one has to find its counterpart 
in the reduced ensembles~:
$${\cal{O}}[A,\bar{\Psi},\Psi] \rightarrow 
{\cal{O}}_1[x,\bar{\xi},\xi] \rightarrow {\cal{O}}_{bp}[T,x] \quad.$$ 
The operators have to fit into the approximation scheme~:
\begin{equation*}
\langle {\cal{O}} \rangle \rightarrow \langle {\cal{O}}_1 \rangle 
\rightarrow \langle {\cal{O}}_{bp} \rangle
\end{equation*}
where the averages are defined as follows~:
\begin{eqnarray*}
\langle {\cal{O}} \rangle  & = & \frac{1}{Z}\  
\int{\cal{D}}A{\cal{D}}\bar\Psi{\cal{D}}\Psi\ 
e^{-{S}[A,\bar\Psi,\Psi]} \ {\cal{O}}[A^{\mu},\bar\Psi,\Psi] \\[0,5cm]
\langle {\cal{O}}_1 \rangle & = &
\frac{1}{Z_1} \int {\cal D} x {\cal D} \bar{\xi} {\cal D} \xi \
e^{-S_1[x,\bar{\xi},\xi]} \
{\cal{O}}_1[x^\mu ,\bar{\xi},\xi] \\[0.5cm]
\langle {\cal{O}}_{bp} \rangle & = & 
\frac{1} {Z_{bp}}\sum_T \int \prod_i{}' {\rm d}^4 x_i  
\ e^{-S_{bp}[T,x]} \ {\cal{O}}_{bp}[T,x]
\end{eqnarray*}

\section{Polyakov-line operator}

The most fundamental operator in the IIB matrix model is the
Polyakov-line operator. The operator is defined as \cite{a1,a3}~:
\begin{equation}
P_k[A] = \frac{1}{N}{\rm Tr} \; e^{ik_{\mu}A^{\mu}},
\end{equation}
where $k$ is some external momentum. We now intend to derive
the form of this operator in the one-loop approximation
$P_{1k}$ and for the branched-polymer ensemble $P_{bp,k}$,
using the approximation scheme described in the previous section~:
$P_k \rightarrow P_{1k} \rightarrow P_{bp,k}$.
As before we shall perform the calcuations 
for classical backgrounds, such that $||x_i-x_j|| \gg \sqrt{g}$.
Finally we will add an effective hard core for small
separations of the classical fields $x_i^\mu$.

In the one-loop calculation we have to expand the operator up to
second order in quantum corrections and 
integrate them out. The first two terms of the expansion are~:
\begin{equation}
P_k[x+a] = 
\frac{1}{N}\left(\sum_{i}e^{ik_{\mu}x^{\mu}_{i}}+\sum_{i<j}
\frac{e^{ik_{\mu}x_{i}^{\mu}}-e^{ik_{\nu}x_{j}^{\nu}}}
{ik_{\rho}(x_{i}^{\rho}-x_{j}^{\rho})} 
(k_{\sigma}a_{ij}^{\sigma})(k_{\tau}a_{ij}^{*\tau})\right) + 
o(a^4) \quad.
\label{pxa}
\end{equation}
This result is derived in the appendix. 
The corrections of order $o(a^4)$ can be neglected within the one-loop approximation. 
The operator is purely bosonic.
The integration
over $\bar{\psi}$, $\psi$ can be carried out indentically to the 
calculations of the partition function \cite{ss}.
Using this result and adding the operator to the integrand
we obtain~:
\begin{equation}
\langle P_k \rangle = {\cal N} \!
\int {\cal{D}}x\ 
[\bigtriangleup^{2}(x)]^{4}{\cal{D}}a{\cal{D}}\xi\  
P_k[x+a]\ e^{-\frac{1}{g^{2}}
\sum_{i<j}||x^{i}-x^{j}||^{2}(\delta_{\mu\nu}+C_{\mu\nu}^{ij})\ 
a^{*\mu}_{ij} a^{\nu}_{ij}}\quad,
\label{intP}
\end{equation}
where ${\bigtriangleup}^{2}(x)\equiv\prod_{i<j}(x_{i}-x_{j})^{2}$. 
The normalization factor ${\cal N}$ is the ratio of the factor,
which arises from the integration over $\bar{\psi}$, $\psi$,
and of $Z$, or alternatively it is a factor which
takes care of the proper normalization $\langle P_{k=0}[x+a] \rangle = 1$. 

We can now insert the first two terms of the expansion $P_k[x+a]$
(\ref{pxa}) into the integral (\ref{intP}). The integrand
is a product of a second order polynomial in the field $a$
and the Gaussian function. Performing the Gaussian integration yields~:
\begin{eqnarray*}
\langle P_{1k} \rangle &=& \frac{1}{Z_1} 
\int {\cal{D}}x{\cal{D}}\bar{\xi} {\cal{D}}\xi\ 
\frac{1}{N}\left(\sum_{i}e^{ik_{\mu}x^{\mu}_{i}}
+g^2 \sum_{i<j} f_{ij} 
k_{\mu}k_{\nu}(\delta^{\mu\nu}+C^{\mu\nu}_{ij})^{-1}\right) 
\\\  &\times&
\prod_{k<l}{\rm \det}_{\mu\nu}^{-1}(\delta^{\mu\nu} +C^{\mu\nu}_{kl})\quad,
\end{eqnarray*}
where the coefficents $f_{ij}$ are given by~:
\begin{equation}
f_{ij} = \frac{e^{ik_{\mu}x_{i}^{\mu}}-e^{ik_{\nu}x_{j}^{\nu}}}
{ik_{\rho}(x_{i}^{\rho}-x_{j}^{\rho}) || x_i-x_j||^2} \ .
\label{fij}
\end{equation}
For a given pair $(ij)$, the matrix $C_{ij}$ (\ref{C}) is antisymmetric
in Lorentz indices $C_{ij}^{\mu\nu}=-C_{ij}^{\nu\mu}$. 
Thus all odd powers of the matrix $C_{ij}$ vanish.
The matrix $C_{ij}$ effectively depends on
one Weyl multiplet $(\bar{\xi}^a_{ij}, \xi^a_{ij})$,
where $\xi^a_{ij}=\xi^a_i-\xi^a_j$. 
For a given pair $(ij)$ 
this multiplet has four independent Grassmannian variables
$(\bar{\xi}^a_{ij}, \xi^a_{ij})$, $a=1,2$.
Any higher than the second power of the matrix $C_{ij}$ 
contains more than four members of the multiplet and therefore
vanishes, since at least one of the Grassmannian variables
enters the product more than once. 
The only non-trivial contribution comes from the
second power $C^2_{ij}$. We have
${\rm tr} \ln (\mathbbm{1}+C_{ij}) = \frac{1}{2} {\rm tr}\; C^2_{ij}$
\, and $k_{\mu}k_{\nu}(\delta^{\mu\nu}+C^{\mu\nu}_{ij})^{-1} =
k_{\mu}k_{\nu}(\delta^{\mu\nu}+(C^2)^{\mu\nu}_{ij})$.
If we insert this to the last equation we obtain~:
\begin{equation}
\langle P_{1k} \rangle = \frac{1}{Z_1} 
\int {\cal{D}}x {\cal{D}}\bar{\xi} {\cal{D}}\xi\ 
P_1[x,\bar{\xi},\xi] \ e^{-S_1[x,\bar{\xi},\xi]}\quad,
\label{p1k}
\end{equation}
where the one-loop action is given by (\ref{S1L})
and the Polyakov-line operator by~:
\begin{equation}
P_1[x,\bar{\xi},\xi] =
\frac{1}{N}\left(\sum_{i}e^{ik_{\mu}x^{\mu}_{i}}
+ g^2 k^2 \sum_{i<j} f_{ij} 
+ g^2 \sum_{i<j} f_{ij} k_\mu\left(C^2_{ij}\right)^{\mu\nu} k_\nu
\right) \, .
\label{p1ko}
\end{equation}
As one can see, the effective Polyakov-line operator,
which in the original model depends only on bosonic fields, 
depends in the one-loop approximation also
on the fermionic zero modes $\bar{\xi},\xi$. 
The dependence is hidden in the matrix $C_{ij}$ (\ref{C}).

The last step is to integrate out the fermionic zero 
modes $\bar{\xi}$ and $\xi$ from the expression (\ref{p1k})
with the operator given by (\ref{p1ko}). The first two terms
of (\ref{p1ko}) do not depend on $\bar{\xi}$ and $\xi$.
Let us integrate  $\bar{\xi}$ and $\xi$ for these two terms.
The only dependence on $\bar{\xi}$ and $\xi$ is now in
the action. Expanding $e^{-S_1[x,\bar{\xi},\xi]}$ in the
integrand of (\ref{p1k}),
we obtain a sum of products over links of graphs $G$
constructed out of links $(ij)$ joining all pairs of 
vertices $x_i$ and $x_j$, $i,j=1,\dots N$~:
\begin{equation}
\exp \frac{1}{2} \sum_{i<j} {\rm tr} \; C^2_{ij} = 
\prod_{i<j} (1 + \frac{1}{2} {\rm tr} \; C^2_{ij}) = 
\sum_{G} \prod_{(ij) \in G} \frac{1}{2} {\rm tr} \; C^2_{ij} \quad.
\label{graphs}
\end{equation}
Each link contributes an entire Weyl multiplet 
$(\bar{\xi}_{ij},\xi_{ij})$
to the product. It is easy to see what happens 
if one integrates a product (\ref{graphs}) 
for the given graph $G$ with the 
measure ${\cal D} \bar{\xi} {\cal D} \xi$. 
The measure contains $N\!-\!1$ independent Weyl multiplets. 
Therefore all integrals for graphs, which have the number of links
different than $N\!-\!1$ vanish. Among the diagrams with $N\!-\!1$
links, only those which have a tree structure survive. To see this,
it is enough to notice that if a graph has a loop
one of the Grassmannian variables enters the integrand twice
and the corresponding product vanishes.
Graphs which have $N\!-\!1$ links and visit all $N$ vertices
are called maximal trees. Thus the integration over zero modes
cuts-out only products of $\frac{1}{2} {\rm tr} \; C^2_{ij}$ 
for $(ij)$ being links of a maximal tree.
When integrating over the $(N\!-\!1)$ independent multiplets
$\bar\xi_i, \xi_i$ for the given maximal tree it
is convenient to change the integration variables to
$\bar{\xi}_{ij}, \xi_{ij}$, where $(ij)$ 
are the links of the tree. The Jacobian 
of the variable change is equal one. 
The integration 
$\prod_{(ij)} {\rm d} \bar{\xi}_{ij} {\rm d} \xi_{ij}$
factorizes in the link variables and can be done independently for
each link of the tree resulting in the following form~:
\begin{equation}
W_{ij}=\frac{1}{2}\int 
{\rm d} \bar{\xi}_{ij} {\rm d} \xi_{ij}\ \bar\xi_{ij}
\Gamma^{\mu\alpha\nu}\xi_{ij}\cdot\bar\xi_{ij}
\Gamma^{\nu\beta\mu}\xi_{ij}\ 
\frac{x_{ij}^{\alpha} x_{ij}^{\beta}}{||x_{ij}||^{8}}\quad,
\label{wij}
\end{equation}
where $x_{ij}=x_i-x_j$.
The integration gives the product over $W_{ij}$ over all
links of the tree. 

One can similarly integrate the third term of 
$P_1[x,\bar{\xi},\xi]$ (\ref{p1ko}). The term
$f_{ij} k_\mu (C^2_{ij})^{\mu\nu} k_\nu$ 
adds a link to each graph in the expansion (\ref{graphs}).
As before the integration over fermionic zero modes
selects only maximal trees, the integrals over link variables factorize and
can be done for each link separately. The new link, which
comes from the third term of the operator (\ref{p1ko})
has now a slightly different fermionic dressing.
The integration of $f_{ij} k_\mu (C^2_{ij})^{\mu\nu} k_\nu$ 
for this link gives~:
\begin{equation}
U_{ij}= f_{ij} \int {\rm d}  \bar{\xi}_{ij} {\rm d}  \xi_{ij} \
\bar\xi_{ij}
\Gamma^{\mu\alpha\nu}\xi_{ij}\cdot\bar\xi_{ij}
\Gamma^{\nu\beta\sigma}\xi_{ij}\ 
\frac{x_{ij}^{\alpha}
x_{ij}^{\beta}k^{\mu}k^{\sigma}}{||x_{ij}||^{8}}\quad,
\end{equation}
which differs from the factors $W_{ij}$ (\ref{wij}) of 
all remaining links of the tree.
For the entire tree $T$ the integration of the third term
gives~:
\begin{equation}
\sum_{(ab)\in T}\left( U_{ab} \prod_{(ij) \ne (ab)} W_{ij} \right)
\quad,
\label{uw}
\end{equation}
where the sum goes over all links of $T$, and the product
over all links of the tree except $(ab)$. 
The link weights $W_{ij}$ and $U_{ij}$ are calculated in
the appendix. The result reads~:
\begin{eqnarray*}
W_{ij} &=& \frac{6}{||x_i - x_j||^6} \\[1cm]
U_{ij} &=& 4 f_{ij} 
\frac{||x_i - x_j||^2 k^2 - \left((x_i-x_j)\cdot k\right)^2}
{|| x_i-x_j||^8}
\end{eqnarray*}
The expression (\ref{uw}) can be rewritten in a form~:
\begin{equation*}
\left( \prod_{(ij)} W_{ij} \right) \sum_{(ab)} \frac{U_{ab}}{W_{ab}} =
\left( \prod_{(ij)} W_{ij} \right) \sum_{(ab)} 
f_{ab} \frac{2( x_{ab}^{2}k^{2}-(x_{ab}k)^{2})}{3|| x_{ab}||^{2}}
\end{equation*}
in which the product of $W_{ij}$ runs over all links
exactly as in the expression for the weight of the tree.

Putting all together, we can write down the result of
the integration of the fermionic zero modes as follows~:
\begin{equation*}
\langle P_{bp,k} \rangle = 
\frac{1}{Z_{bp}}\sum_{T} \int\prod_{i}{}' {\rm d}^4 x_{i} 
\left( \prod_{(ij)\in T} W_{ij}\right) \ P_{bp,k}[x]\quad,
\end{equation*}
where the operator is given by
\begin{equation}
P_{bp,k}[x] = 
\frac{1}{N}\left\{\sum_{a}e^{ik_{\mu}x^{\mu}_{a}}+
g^2 k^{2}\sum_{a<b} f_{ab}+ g^2 \! \sum_{(ab)\epsilon{T}}
f_{ab}\frac{2( x_{ab}^{2}k^{2}-
( x_{ab}k)^{2})}{3|| x_{ab}||^{2}}\right\}
\label{pbp}
\end{equation}
Everything in this section was done under the assumption that
$||x_i-x_j|| \gg \sqrt{g} $. The product of $W_{ij}$
describes the weight of the model for large distances.
It is equivalent to the effective action 
$S_{bp} =  6 \sum_{(ij)\in T} \ln || x_i -x_j||$
for large distances between $x_i$'s.
For small distances, we have to introduce an effective
repulsion core as described in the previous section, eventually obtaining the effective action (\ref{s1bp}).
The Polyakov-line operator is given in the branched-polymer
model by (\ref{pbp}).

\section{Discussion}
It is quite surprising that the partition function (\ref{Z}),
a model defined by
supersymmetric Yang-Mills integrals, can be approximated 
by the partition function of a very simply statistical model 
of branched-polymers. 
However, as we demonstrated in this paper, observables which 
are given by simple expressions in the supersymmetric 
Yang-Mills model become rather complicated in the corresponding
statistical model of branched-polymers. With the explicit 
form of the Polyakov-line operator, which was derived 
analytically in this paper in the one-loop approximation,
one should be able in the future to study numerically 
its scaling properties and to discuss limits of 
applicability of the one-loop approximation.

\section*{Acknowledgments}

This work was supported by Polish State 
Committee for Scientific
Research (KBN), grant 2P03B 09622 (2002-2004) and by the EC IHP network
HPRN-CT-1999-000161. M.W. thanks the University of Bielefeld for a graduate
scholarship during the first stage of this work. We thank Romuald Janik and J\"org
Erdmann for helpful discusions. 

\section*{Appendix A}
In this part of the appendix, we calculate the second order terms in 
off-diagonal quantum
corrections $a_{ij}$ (\ref{xa}) to the Polyakov-line operator
(\ref{pxa}). 
\begin{equation}
P_{k}[A]=\frac{1}{N}{\rm Tr} \; e^{ik_\mu A^\mu} =
\frac{1}{N}{\rm Tr}\; e^{\chi + \alpha} \quad,
\label{ca}
\end{equation}
where the matrix $\chi_{ij} = \chi_i \delta_{ij} =
i k_\mu x^\mu_i \delta_{ij}$ is 
diagonal and $\alpha_{ij} = i k_\mu a^\mu_{ij}$ is off-diagonal. 
The zeroth order term of the expansion in $\alpha$ of
the right hand side of (\ref{ca}) is
$\frac{1}{N} \sum_i e^{ik_\mu x^\mu_i}$, while the first order vanishes.
The second order terms can be collected to the expression~:
\begin{equation}
\sum_{n=2}^{\infty} 
\frac{1}{n!} \sum_{r+q=n-2} (r+1)\ {\rm Tr} \;
\left( \chi^{r} \alpha \chi^{q} \alpha \right)\quad.
\label{sn}
\end{equation}
Because $\chi$ is diagonal and $\alpha$ is off-diagonal
we can write~:
\begin{equation*}
{\rm Tr} \; \left( \chi^{r} \alpha \chi^{q} \alpha \right)=
\sum_{ij} \chi_{i}^r \chi_{j}^{q} \alpha_{ij} \alpha_{ji}=
\sum_{i<j} (\chi_{i}^{r} \chi_{j}^{q} +\chi_{i}^{q} \chi_{j}^{r}) 
\alpha_{ij} \alpha_{ji}\quad.
\end{equation*}
We thus have to evaluate
\begin{equation*}
\sum_{r+q=n-2}  (r+1) (\chi_i^r \chi_j^q+\chi_i^q \chi_j^r)\quad.
\end{equation*}
This can be rewritten as
\begin{equation*}
\left( \frac{\partial}{\partial \chi_i} \chi_i + 
\frac{\partial}{\partial \chi_j}\chi_j \right)
\left[ \sum_{r+q=n-2} \chi_i^r \chi_j^q \right]\nonumber\quad,
\end{equation*}
while the term in square brackets is just
\begin{equation*}
\frac{\chi_i^{n-1}-\chi_j^{n-1}}{\chi_i-\chi_j}\quad.
\end{equation*}
Acting on it with the differential operator we get
\begin{equation*}
\left( \frac{\partial}{\partial \chi_i} \chi_i + 
\frac{\partial}{\partial \chi_j}\chi_j \right)
\left[\frac{\chi_i^{n-1}-\chi_j^{n-1}}{\chi_i-\chi_j}\right]=
\frac{n \chi_i^{n-1}-n \chi_j^{n-1}}{\chi_i-\chi_j}\quad.
\end{equation*}
Performing the summation over 
$n$ as in the equation (\ref{sn}) we can write the
sum of the second order terms in a compact form~:
\begin{equation*}
\sum_{i<j}\frac{e^{\chi_i} - e^{\chi_j}}{\chi_i-\chi_j} 
\alpha_{ij}\alpha_{ji} =
\sum_{i<j} 
\frac{e^{ik_{\mu}x_{i}^{\mu}}-e^{ik_{\nu}x_{j}^{\nu}}}
{ik_{\rho}(x_{i}^{\rho}-x_{j}^{\rho})} 
k_{\sigma}k_{\tau}a_{ij}^{\sigma}a_{ji}^{\tau} \quad,
\end{equation*}
which leads to the equation (\ref{pxa}).

\section*{Appendix B}
We present here an explicit calculation 
of the link-weights $W_{ij}$ and $U_{ij}$.
Throughout this appendix, we denote the members of the
multiplet $(\bar{\xi}_{ij},\xi_{ij})$ by 
$(\bar{\xi},\xi)$,
to simplify the notation. This does not lead to a confusion, 
because the integrals $W_{ij}$ and $U_{ij}$ are 
one-link integrals and involve only integration over one 
Weyl multiplet.

From this point of view $\xi_{ij}$ plays merely
the role of the integration variable and can be renamed.

For a Weyl spinor $\bar\xi_a,\xi_a$, $a=1,2$ in $D=4$ 
we have
\begin{equation*}
\int {\rm d}\bar\xi_{1} {\rm d}\bar\xi_{2}
{\rm d}\xi_{1} {\rm d}\xi_{2}\ 
\bar\xi_{a}\xi_{b}\bar\xi_{c}\xi_{d}=
-\epsilon_{ac}\epsilon_{bd}=
\delta_{ad}\delta_{cb}-\delta_{ab}\delta_{cd}\quad.
\end{equation*}
In the Weyl representation, the gamma matrices are given by
\begin{equation*}
\Gamma^{k}=\left(\begin{array}{cc}
0&\sigma_{k}\\
\sigma_{k}&0\\
\end{array}
\right), \
\Gamma^{4}=\left(\begin{array}{cc}
0&-i\mathbbm{1}\\
i\mathbbm{1}&0\\
\end{array}
\right) , \
\Gamma_{5}=\left(\begin{array}{cc}
\mathbbm{1}&0\\
0&-\mathbbm{1}\\
\end{array}
\right)\quad,
\end{equation*}
where $\sigma_{k}, k=1,2,3$ are the Pauli matrices.
We can write 
\begin{equation*}
\Gamma^{\mu}=\left(\begin{array}{cc}
0&\Gamma^{\mu}_{-}\\
\Gamma^{\mu}_{+}&0\\
\end{array}
\right) \ , \quad
\Gamma^{\mu}\Gamma_{5}=\left(\begin{array}{cc}
0&-\Gamma^{\mu}_{-}\\
+\Gamma^{\mu}_{+}&0\\
\end{array}
\right)\quad.
\end{equation*}
We will need a second expression in the calculation
involving the symbol~:
$$\Gamma^{\mu\alpha\nu}=
\epsilon^{\mu\alpha\nu\sigma}\Gamma^{\sigma}\Gamma_{5}\quad.$$

\noindent
Now we can calculate $W_{ij}$~:
\begin{eqnarray*}
W_{ij}&=& 
\frac{1}{2}\int {\rm d}\bar\xi_{1} {\rm d}\bar\xi_{2}
{\rm d} \xi_{1} {\rm d} \xi_{2}\quad
\bar\xi\Gamma^{\mu\alpha\nu}_{-}
\xi\cdot\bar\xi\Gamma^{\mu\beta\nu}_{-}\xi\ \frac{
x_{ij}^{\alpha}x_{ij}^{\beta}}{|| x_{ij}||^{8}}\\[0.3cm]
&=&-\frac{1}{2}\epsilon_{ac}\epsilon_{bd}\epsilon^{\mu\alpha\nu\sigma}
\epsilon^{\mu\beta\nu\tau}\Gamma^{\sigma}_{-ab}\Gamma^{\tau}_{-cd}\
\frac{
  x_{ij}^{\alpha} x_{ij}^{\beta}}{|| x_{ij}||^{8}}\\[0.3cm]
&=&\frac{1}{2}\underbrace{(\Gamma^{\sigma}_{-ab}\Gamma^{\tau}_{-ba}-
\Gamma^{\sigma}_{-cc}\Gamma^{\tau}_{-dd})}_{2\delta^{\sigma\tau}}
\epsilon^{\mu\alpha\nu\sigma}\epsilon^{\mu\beta\nu\tau}\
\frac{
 x_{ij}^{\alpha} x_{ij}^{\beta}}{|| x_{ij}||^{8}}\\[0.3cm]
&=&
\underbrace{\epsilon^{\mu\alpha\nu\sigma}
\epsilon^{\mu\beta\nu\sigma}}_{3!\delta^{\alpha\beta}}
\frac{
  x_{ij}^{\alpha} x_{ij}^{\beta}}{|| x_{ij}||^{8}} \\[0.3cm]
&=&\frac{6}{|| x_{ij}||^{6}}
\end{eqnarray*}
Similiarly for $U_{ij}$
\begin{eqnarray*}
U_{ij}&=&f_{ij}\int {\rm d}\bar\xi_{1} {\rm d}\bar\xi_{2}
{\rm d}\xi_{1} {\rm d} \xi_{2}\quad
\bar\xi\Gamma^{\mu\alpha\nu}_{-}\xi\cdot\bar\xi
\Gamma^{\sigma\beta\nu}_{-}\xi\ 
\frac{ x_{ij}^{\alpha} x_{ij}^{\beta}
k^{\mu}k^{\sigma}}{|| x_{ij}||^{8}}\\[0.3cm]
&=&-\epsilon_{ac}\epsilon_{bd}\epsilon^{\mu\alpha\nu\rho}
\epsilon^{\sigma\beta\nu\tau}\Gamma^{\rho}_{-ab}\Gamma^{\tau}_{-cd}\
 f_{ij}\frac{ x_{ij}^{\alpha} x_{ij}^{\beta}
k^{\mu}k^{\sigma}}{|| x_{ij}||^{8}}\\[0.3cm]
&=&\underbrace{(\Gamma^{\rho}_{-ab}\Gamma^{\tau}_{-ba}-
\Gamma^{\rho}_{-cc}\Gamma^{\tau}_{-dd})}_{2\delta^{\rho\tau}}
\epsilon^{\mu\alpha\nu\rho}\epsilon^{\sigma\beta\nu\tau}\
 f_{ij}\frac{
  x_{ij}^{\alpha} x_{ij}^{\beta}
k^{\mu}k^{\sigma}}{|| x_{ij}||^{8}}\\[0.3cm]
&=& 4 f_{ij}(\delta^{\mu\sigma}\delta^{\alpha\beta}
-\delta^{\mu\beta}\delta^{\alpha\sigma})
\frac{ x_{ij}^{\alpha} x_{ij}^{\beta}
k^{\mu}k^{\sigma}}{||x_{ij}||^{8}}\\[0.3cm]
&=&
4 f_{ij}\frac{x_{ij}^{2}k^{2}-( x_{ij}k)^{2}}
{|| x_{ij}||^{8}}
\end{eqnarray*}

\end{document}